\documentclass[english,floatfix,twocolumn,showpacs,prb]{revtex4}

\usepackage[T1]{fontenc} 
\usepackage[latin1]{inputenc}
\usepackage{graphicx} 
\usepackage{dcolumn} 
\usepackage{bm} 
\usepackage{ae,aecompl}
 
\begin{document} 

\title{The role of nitrogen related defects in high-$k$ dielectric oxides:
  density functional studies}

\author{J. L. Gavartin$^1$,
A. S. Foster$^2$, G. I. Bersuker$^3$ and A. L. Shluger$^1$} \affiliation{ $^1$Department of Physics and
Astronomy, University College London, Gower Street, London WC1E 6BT,
UK. \\ $^2$Laboratory of Physics, Helsinki University of Technology,
P.O. Box 1100, FIN-02015 HUT,  Finland \\ $^3$International Sematech,
Austin, TX 78741, USA.  } \date{\today}

\begin{abstract}
Using \emph{ab initio} density functional total energy and molecular dynamics 
simulations, we study the effects of various forms of nitrogen post deposition 
anneal (PDA) on the electric properties of hafnia in the context of
its application as a gate dielectric in field effect transistors (FET).
We consider the atomic structure and energetics of nitrogen containing defects 
which can be formed during the PDA in various N-based ambients: 
N$_2$, N$^+_2$, N, NH$_3$, NO, N$_2$O. We analyse the role of such defects  
in fixed charge accumulation, electron trapping and in the growth of the 
interface SiO$_2$ layer. We find that nitrogen anneal of the oxides leads 
to an effective immobilization of native defects such as oxygen
vacancies and interstitial oxygen ions, which may inhibit growth of silica
layer. Nitrogen in any form effectively incorporates into the pre-existing
oxygen vacancies and, therefore may decrease the concentration of shallow
electron traps. However, nitrogen in any form is unlikely to significantly 
reduce the fixed charge in the dielectric. 
\end{abstract}
 

\maketitle

\section{Introduction}
Transistor scaling, which enables continuous increase of performance
of integrated circuits, has been generally facilitated by a reduction
in the thickness of the gate dielectric in typical metal oxide
semiconductor field effect transistors (MOSFET). However, a further 
decrease of the thickness of conventional silicon dioxide (SiO$_2$)-type 
dielectrics leads to an unacceptable rise in the gate  leakage current, 
and hence, power consumption. Several
possible solutions are under consideration \cite{highk_nature}, but
one of the most attractive, which retains standard MOSFET design, is to replace
SiO$_2$ with a material of higher dielectric constant (high-$k$). The
resulting increase in effective capacitance means that a thicker gate
dielectric layer can be used, reducing gate leakage current, while
providing comparable performance to a much thinner SiO$_2$ layer. The
choice of specific high-$k$ material is subject to rather strict electrical and 
process
integration requirements \cite{Huff_2003} eliminating most of the potential 
candidates, and presently hafnium oxide-based dielectrics are considered to 
be the most promising from a practical standpoint.

In spite of recently reported successes \cite{Chau_2003}, the performance of
high-$k$ transistors requires  further improvement to meet the industry
needs for most of the potential applications. Most critically, high-$k$
devices suffer from poor channel mobility
\cite{Wilk_review,Wong_ibm2002rev,Young_1_2004,Young_2_2004}  and instabilities of the threshold
potential (V$_T$) \cite{Kerber_ieee2003_1,Carter_aps2003}. It is understood that  these
drawbacks result from the high density of structural defects in the
high-$k$ films \cite{Bersuker_mat2day_2004}, in particular, defects acting as
electron  traps 
\cite{Stesmans_Afanasev_esr_book_2004,Afanasev_apl2002,Haussa_jecs2002}.
These defects were shown to form localized states in the band gap, whose
occupancy may vary with the external chemical potential 
\cite{Foster_2002_2,foster02_1,nano_book_2003}. It was generally expected that a 
post deposition anneal
(PDA) would reduce the density of intrinsic defects. However, for high-$k$ films
deposited on silicon, PDA usually results in a decrease of the effective 
capacitance of the gate stack. This decrease was attributed to a spontaneous 
growth of the SiO$_x$ buffer layer between high-$k$ and silicon substrate during 
the high-$k$ film deposition and subsequent anneals (600 - 1000 C), facilitated 
by the oxygen excess. The low quality of the resulting interfacial SiO$_x$ 
layer negatively impacts transistor characteristics  \cite{Bersuker_jjap_2004}. 
In order to suppress the suboxide growth, nitrogen 
rich ambients are preferable in the PDA \cite{Wilk_review}. In general, nitrogen 
incorporation into the Hf-based dielectrics was shown to provide several 
advantages - reducing boron penetration from the p-type poly silicon gate 
electrodes, improving electrical performance and increasing stability of high-$k$ devices. 
Besides N$_2$ PDA, several other options were used to increase nitrogen 
incorporation in Hf-based oxides, including remote nitrogen plasma 
(N$_2^+$) \cite{Lukovsky_ass2003}, ammonia (NH$_3$), nitrogen monoxide (NO), 
and N$_2$O, as well as nitrato-CVD 
deposition processes \cite{Afanasev_jap2004,afanasev04_offs}. However, the lack of understanding of 
forms of nitrogen incorporation and its effect on the electronic properties of 
the dielectric hampers further progress.

In this paper we theoretically investigate the incorporation of N$_2$, N, NO, 
NH$_3$ species in different charge states into the ideal and defective 
HfO$_2$ crystal. Specifics of the structural and chemical 
composition of real films depend strongly on the deposition and PDA conditions. 
In particular, a thin film deposited at relatively low temperatures (typically 
below 500 K) is in a metastable porous low density state, often characterized by 
substantial non-stoichiometry. Significant efforts have been recently made 
to resolve the atomic structure of the deposited films 
\cite{Stemmer_nano_2003,Opila_Eng_2002}.  After PDA in oxygen (T > 770 K), 
hafnia films on Si adopt the monoclinic structure \cite{hf_gap,hf_ald2,hf_csd}. 
This is in contrast with zirconia which may crystallize into either monoclinic 
or tetragonal structure  depending on the film thickness and stoichiometry
\cite{Stemmer_nano_2003}. In this paper we leave the problem of 
polymorphism aside, and focus on monoclinic hafnia. 

In broad terms, an optimal post deposition anneal should stipulate: 
\begin{itemize}
\item formation
of dense (preferably amorphous) oxide films with controlled thickness, 
high dielectric constant and high thermal stability; 
\item reduction 
of fixed charge and electron traps in the bulk of the film and at the interface 
with Si (high electrical stability) 
\item an effective control over the growth of 
the SiO$_2$ buffer layer; 
\item inhibition of the diffusion of charged species 
towards the interface and into the channel region.
\end{itemize} 

In order to assess the effectiveness of different anneals, we assume the following
physical picture. After deposition and PDA, a hafnia film is crystallized into 
the monoclinic phase which may still contain a significant concentration of 
oxygen vacancies and interstitials. 
In the significantly non-stoichiometric films
either oxygen interstitials (oxygen excess) or oxygen vacancies (oxygen depletion) 
will dominate. Nitrogenous species introduced by the PDA, diffuse into the bulk 
oxide and take part in the following (not necessarily sequential) processes: 
{\em i)} incorporation in the lattice interstitials or replacement of lattice 
oxygen ions; 
{\em ii)} dissociation into other species both in interstitial and in regular lattice 
sites; 
{\em iii)} reaction with the existing oxygen vacancies and interstitial oxygen species.

\begin{figure}
\includegraphics[scale=0.50]{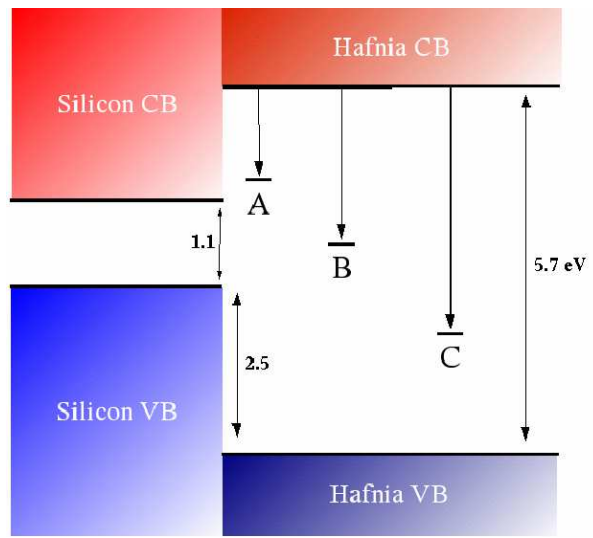}
\caption{Schematic figure showing classes of defect levels with
  respect to hafnia and silicon bands. The band offset for silicon
  is taken from ref. \cite{afanasev02_offs,afanasev04_offs}.
\label{defects_scheme}}
\end{figure}

The nitrogen defects thus formed may be neutral with respect to the lattice or 
themselves form fixed charge or charge trap centres. This would depend on the 
position of the corresponding defect levels with respect to the band edges of 
Hafnia and Silicon and with respect to the Fermi-level 
(see Fig. \ref{defects_scheme}). The latter in the n-channel devices is 
located near the silicon conduction band minimum (CBM). 


All the electrically active defects in the oxide layer can be qualitatively 
classified by the position of their levels with 
respect to the silicon band edges:

{\it A)} Defects whose levels are resonant with the Si conduction band. 
Under zero electric field and in the thermodynamic equilibrium these 
levels are empty. However, such states are available for resonant tunneling 
at non-zero gate voltage, and thus may serve as electron traps. 

{\it B)} Defects whose levels fall into the silicon band gap. The electron
occupancy of such defects depends on the position of the Fermi level. 
These deep defects are responsible for the Fermi-level pinning, and thus,
for the high threshold voltages. Also they may contribute into threshold 
potential instability and, to some extent, to a fixed charge problem, since
the charge relaxation time on such defects may be macroscopically slow. 

{\it C)} Defect states resonant with the silicon valence 
band are expected to be occupied, and, depending on their charge 
with respect to the lattice, they may become a major source of the fixed 
charge.

Therefore, the efficiency of a specific PDA could be judged by the integral 
effect of the resultant nitrogen contained species on the bulk charge trap 
density, and on the mobility of oxygen, which may occur via the vacancy or 
interstitial mechanisms \cite{foster02_1}. 
 
Although the model outlined above is certainly simplified, it will form 
a framework for systematizing the effect of various nitrogenous species on the 
properties of monoclinic hafnia. It can be then refined to include other 
important effects e.g. interface diffusion and segregation, impurity 
clustering etc.

In the next section we give the details of the calculations. The results of the 
modeling of the effect of different anneal gases are presented in section III 
and discussion is given in the last section.

\section{Calculation procedure}

All the calculations were performed using the VASP code \cite{VASP1,VASP2}, 
implementing spin-polarized Density Functional Theory (DFT) and the Generalized 
Gradient Approximation of Perdew and Wang \cite{GGA-PerdewII} known as GGA-II. 
The plane wave basis set was used in conjunction with the ultra-soft 
pseudopotentials of Vanderbilt type \cite{pseudo1}.  The standard 
pseudopotentials for hydrogen, hafnium, nitrogen, and oxygen atomic cores were 
generated with charges +1,+4,+5, and +6  respectively \cite{pseudo2}.

Bulk hafnia at atmospheric pressure and low temperature has monoclinic 
symmetry \cite{hf_basic_struc1,hf_basic_struc2} (space group $P2_{1}/c$). At T = 
2000 K it undergoes martensitic phase transformation into the tetragonal phase 
(space symmetry $P4_2/nmc$), and above T = 2870 K the cubic fluorite structure 
is most stable (space symmetry $Fm3m$). The monoclinic structure is 
characterized by the two non-equivalent anion sublattices (Fig. 
\ref{n2_struct}). In one, oxygen ions are 4-fold coordinated, and in another 
they are 3-fold  coordinated. The equilibrium O-Hf distances range between 2.14 
and 2.24 \AA\ in the former and between 2.05 and 2.14 in the latter sublattice. 
These structural parameters are well reproduced using DFT and pseudopotentials 
described above and are discussed in detail in ref \cite{nano_book_2003}.

The calculations were made using a 96 atom unit cell, which is a 
$2\times2\times2$ extension of the 12 atom monoclinic primitive unit cell. We 
used two $k$-points in the  irreducible part of the Brillouin zone and 
a cutoff energy of 400 eV for atomic relaxation and total energy calculations. 
The total energies of the charged systems were corrected for the spurious 
electrostatic interaction arising from the periodic boundary conditions 
\cite{Leslie-Gillan,lev_coul}.

To test the stability of atomic configurations optimized in static calculations 
and to extract dynamical information, such as diffusion paths and 
vibrational frequencies, we carried out an extensive set of Born-Oppenheimer 
molecular dynamics (MD) simulations using VASP. In these simulations 
we used $k$=0 and 300 eV energy cutoff and a constant volume, 
energy and number of particles (NVE) algorithm with a time step of 1 fs. This 
assured a total energy drift of less than 0.05 eV at T = 600 K for a typical MD run 
of 5 ps. The supercell volume in all calculations was fixed to its 0 K 
theoretical value \cite{nano_book_2003}. 

For each charge state $q$ of a candidate defect ($D$) we first determine its 
geometric structure and formation or incorporation energy ($G(D^q)$) using 
the static DFT calculations. Given the macroscopically long defect diffusion 
and charge relaxation times \cite{Carter_aps2003}, we believe that the defect 
equilibrium Free energies do not necessarily reflect relative concentrations of 
various defects  (even if the Fermi level and atomic chemical potentials were 
well defined). Thus, we do not discuss free energies, 
but rather assume initially that all energetically stable charge states are possible, 
and then discuss most probable charge states considering there relative electron 
affinities of the defects.
   
The position of defect electrical levels with respect to 
the CBM of hafnia and silicon determines 
the stability of a defect in a particular charge state as well as its
ability to trap or release an electron. These properties are quantified by the
vertical and relaxed defect ionization energies ($I_{p}(D^{q})$), and vertical 
electron ($\chi _{e}(D^{q})$) and hole ($\chi_{h}(D^{q})$) affinities, which are 
calculated using the approach described in ref. \cite{Foster_2002_2} 
(see Appendix for details). This approach is based on standard definitions of 
electron affinities and ionization energies, it employs combinations of total 
energies of the system in different charge states and conserves the number of 
electrons. In order to discuss the electron and hole trapping by defects from 
silicon, we neglect the band bending at the interface and the effect of applied 
voltage and use the experimental band offsets between bulk hafnia and 
silicon \cite{afanasev02_offs,afanasev04_offs}. 

Towards the full picture of possible products of incorporation of 
nitrogen, we examine all possible dissociation channels for nitrogenous molecular 
species in hafnia matrix. This is done by calculating their dissociation energy 
into various products. The latter is found as the difference of total energies for 
the system with the associated defect and the joint energy of the infinitely separated 
dissociation products, corrected for double counting of the perfect supercell 
term. In further discussion we treat a molecular defect as thermodynamically 
unstable if its dissociation is exothermic. 

\section{Results of Calculations}

As discussed previously, nitrogen-based PDA can be performed using a variety 
of gas sources (ambients), and depending on the  chemical composition, pressure 
and temperature, different products may occur in the bulk film. Here we
consider each PDA ambient in turn, and predict the likely resultant
defects and their effects on the properties of the oxide.

\subsection{Molecular Nitrogen} 

\subsubsection{N$_2$ incorporation} 

The experimental dissociation energy of an N$_{2}$ molecule in the gas 
phase is $\sim$9.9 eV  \cite{Kutzler-Painter_prb1988} (10.0 eV in our DFT 
calculations). Thus, no N$_2$ dissociation is likely to occur at the surface, 
and molecular nitrogen is expected to diffuse inside the bulk oxide. 
Indeed, as we show below, even trapping of extra electrons or interaction 
with oxygen vacancies does not lead to a spontaneous dissociation of N$_{2}$ 
molecular species inside the oxide. 

\begin{figure}
\includegraphics[scale=0.50]{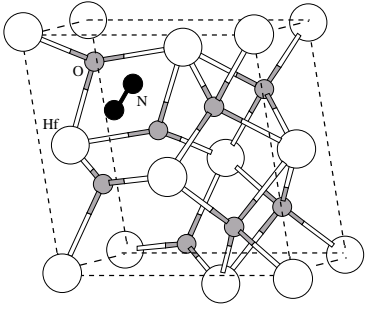}
\caption{Structure of neutral nitrogen interstitial molecule in
hafnia. The ions are colour coded as follows: white - Hf, grey - O, black - N.
\label{n2_struct}}
\end{figure}

\begin{figure}
\includegraphics{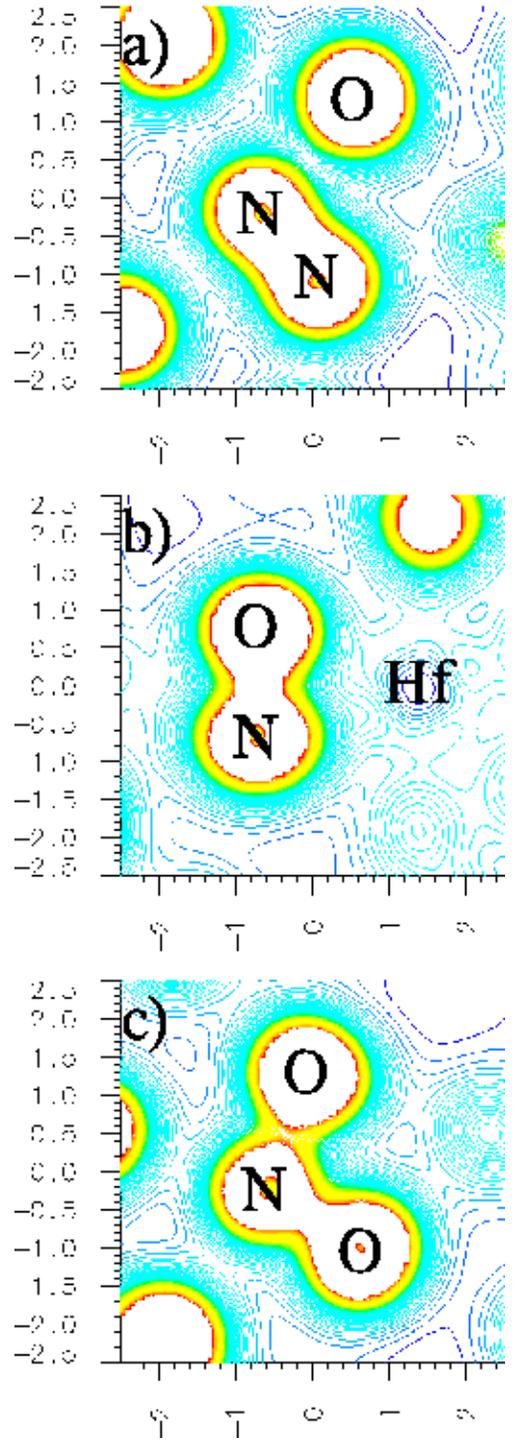}
\caption{Charge density 2D slices around various nitrogen defects in
hafnia: (a) plane through an
interstitial nitrogen molecule and a three coordinated oxygen
atom; (b) plane through a neutral nitrogen interstitial, three
coordinated oxygen and hafnium; (c) plane through a single positively
charged nitrogen interstitial and two lattice oxygens.  \label{charge_dens}}
\end{figure}

The equilibrium configuration of the neutral interstitial nitrogen 
molecule is shown in Fig. \ref{n2_struct}. The incorporation energy for 
this configuration with respect to N$_2$ in the gas phase is $\sim$3.0 eV 
per nitrogen atom. It is seen in figures 
\ref{n2_struct} and \ref{charge_dens}(a) that the interstitial  N$_2$ does not 
form a bond with the lattice oxygen, but rather incorporates in the 
interstitial space near to the three-fold coordinated oxygen site. This is in 
contrast with the O$_2$ molecule, which makes a {\it covalent bond} with the 
lattice anion \cite{Foster_2002_2}.

The interstitial N$_2$ molecule has a positive affinity to two electrons 
(defect electron affinities are given in Table \ref{tab_aff}).  The localization 
of the first extra electron on the N$_{2}$ interstitial (N$^{-}_{2}$ radical)  
results in the outward relaxation of the nearest neighbor oxygen ions (cf. O-N  
distance of 1.96 \AA\ in N$^{0}_{2}$ and $\sim$2.20 \AA\ in N$^{-}_{2}$). The 
corresponding relaxation energy is $\sim$1.2 eV. The second extra electron is 
also strongly localized on the N$^{2-}_{2}$ molecular ion. This further 
increases the O-N distance to 2.34 \AA\ with a relaxation energy of 0.6 eV with 
respect to the N$^{-}_{2}$ geometry. Interestingly, a hole (an absence of the 
electron) in HfO$_2$ will not localize on the N$_{2}$ interstitial, owing to the
fact that the last occupied orbital of the molecule is well below the
valence band maximum (VBM).

\begin{table}[htbp]
\squeezetable
\caption{Vertical ionizational potential $I(D)$, relaxed electron $\chi_{e}(D)$ 
and vertical hole $\chi_{h}(D)$ affinities (in eV) of
defects in different charge states in hafnia.  The subscript {\it O}
indicates the substitution in the 3-coordinated oxygen site.}
\begin{ruledtabular}
\begin{tabular}{lccr}
$D$& $I_{p}(D)$& $\chi_{e}(D)$& $\chi_{h}(D)$
\tabularnewline \hline \hline  N$^{-}$& 5.0& - & 0.7
\tabularnewline  N$^{0}$& 5.3& 4.2& 0.3
\tabularnewline   N$^{+}$& 5.6& 4.1& 0.1
\tabularnewline   N$^{2+}$& -& 5.0& -
\tabularnewline \hline N$_{2}^{2-}$& 4.2& -& 1.5
\tabularnewline   N$_{2}^{-}$& 4.3& 3.4& 1.3 
\tabularnewline   N$_{2}^{0}$& -& 3.3& - 
\tabularnewline \hline   NO$^{0}$& 4.8 & 4.4 & 0.7
\tabularnewline \hline   N$^{-}_{O}$ & 5.2& -& 0.5
\tabularnewline   N$^{0}_{O}$& 5.6& 4.6& 0.1
\tabularnewline   N$^{+}_{O}$& -& 5.2& -
\tabularnewline \hline  (N$_{2}$)$^{-}_{O}$& 3.9& -& 1.7
\tabularnewline (N$_{2}$)$^{0}_{O}$ & 4.4& 3.3& 1.1
\tabularnewline (N$_{2}$)$^{+}_{O}$ & -& 3.8& -
\tabularnewline 
\end{tabular}
\end{ruledtabular}
\label{tab_aff} 
\end{table}

\subsubsection{Thermodynamic considerations on the likely charge states}

We have established that interstitial nitrogen molecules may have multiple charge states,
whose occurance at thermodynamic equilibrium is defined explicitely by the electron 
chemical potential. The latter depends not only on the external potential, but also on
the concentrations and electrical levels of all other defects present in the lattice.
In this case, the problem of the possible defect charge states may be evaluated by
considering relative electron affinities of various defects, and answering the question:
given the fixed number of electrons, on which of the defects they are more likely to 
reside? This question can be resolved by considering the electron exchange reactions
between the pairs of infinitely separated defects as presented in Table \ref{tab_elex}.

\begin{table} 
\squeezetable
\caption{Electron exchange reactions for defects in m-HfO$_2$. The energies 
are calculated in the assumption of infinitely separated defects. $E>0$
indicates exothermic reaction along the arrow.}
\begin{ruledtabular} 
\begin{tabular}{llr} 
No & Reaction & Energy (eV)           \\ \hline   
1 & $ O^0 + V^0   \Rightarrow O^- + V^+           $  &  1.1   \\  
2 & $ O^0 + V^0   \Rightarrow O^{--} + V^{++}     $  &  2.4   \\  
3 & $ N_2^0 + V^0  \Rightarrow N_2^- + V^+       $  &  0.5   \\  
4 & $ N_2^0 + V^0  \Rightarrow N_2^{--} + V^{++} $  &  0.7   \\  
5 & $ N_2^0 + V^+  \Rightarrow N_2^{-} + V^{++}  $  &  0.1   \\  
6 & $ N_2^- + V^0  \Rightarrow N_2^{--} + V^{+}  $  &  0.6   \\  
7 & $ N^+ + V^0   \Rightarrow N^0 +    V^+        $  &  1.3   \\  
8 & $ N^0 + V^0   \Rightarrow N^- +    V^+        $  &  1.4   \\  
9 & $ NH_2^+ + V^0  \Rightarrow NH_2^-  + V^{++} $  &  1.5   \\  
10 & $ H^+    + V^0  \Rightarrow  H^-  + V^{++}   $  & -0.4   \\ 
\hline  
11 & $ N_2^- + O^-  \Rightarrow N_2^{--}  + O^0  $  & -0.5   \\ 
12 & $ N_2^-    + O^-  \Rightarrow N_2^0  + O^{--} $  & 1.2  \\ 
13 & $ N_2^{--} + O^0  \Rightarrow N_2^0  + O^{--} $  & 1.7  \\ 
14 & $ N^0    + O^0  \Rightarrow N^+  + O^-       $  &  -0.2  \\ 
15 & $ N^0   + O^-  \Rightarrow N^-  + O^0       $  &  0.3   \\ 
16 & $ N^-    + O^-  \Rightarrow N^0  + O^{--}   $  &  0.3   \\ 
17 & $ N^{++}    + O^{--}  \Rightarrow N^+  + O^{-}   $  &  0.6   \\ 
18 & $ N^{++}    + O^{-}  \Rightarrow N^+  + O^{0}   $  &  1.2   \\ 
\hline
19 & $ 2V^{+}  \Rightarrow V^0  + V^{++}         $   & 0.3  \\ 
20 & $ 2O^{-}  \Rightarrow O^0  + O^{--}         $   & 0.6  \\ 
21 & $ 2N_2^{-}  \Rightarrow N_2^0  + N_2^{--}   $   & 0.1  \\ 
22 & $ 2N^{0}  \Rightarrow N^+  + N^{-}          $   & 0.1  \\ 
23 & $ 2NH_2^{0}  \Rightarrow NH_2^+  + NH_2^{-} $   & 2.8  \\ 
24 & $ 2H^{0}  \Rightarrow H^+  +  H^-           $   & 2.7  \\ 
\end{tabular} 
\end{ruledtabular} 
\label{tab_elex} 
\end{table} 

Consider for example the oxygen Frenkel pair (an interstitial atom and the 
vacancy).
It is evident from reactions (\ref{tab_elex}.1,2) that in the overall neutral 
m-HfO$_2$ the zero temperature equilibrium state corresponds to the doubly 
positively charged vacancies and doubly negatively charged interstitial oxygen 
atoms. Next, consider N$_2$ interstitial molecules. Reactions 
(\ref{tab_elex}.3-6) suggest that they are more 
electronegative then anion vacancies. Therefore, in oxygen deficient films (high 
concentration of vacancies) nitrogen molecules will be predominantly doubly negative, 
while the vacancies remain doubly positive. 

In the case of oxygen excess (large concentrations of interstitials), the relative charge 
state is defined by the energy balance in the reactions (\ref{tab_elex}.11-13). It 
follows that interstitial oxygen ions are more electronegative than nitrogen molecules. 
However, as discussed above, N$_{2}^{+}$ is not stable, so nitrogen interstitial 
molecules will remain largely neutral in this case.

Finally, reaction (\ref{tab_elex}.21) suggests that neutral and doubly negative
N$_2$ interstitials are only marginally preferable over the N$_2^-$.

The energetics of the electron exchange reactions considered in Tab. \ref{tab_elex} 
allows to narrow down the range of relevant association/dissociation reactions between 
various pairs of defects considered below. 

\subsubsection{N$_2$ reactions with interstitial oxygen species}

First, we observe that the interstitial oxygen 
molecules (O$_2$) are either marginally stable or unstable in any charge state 
(Table \ref{tab_NO}.1-4). Therefore, we conclude  
that excess oxygen is present only in the form of atomic interstitials. 
As demonstrated in
refs. \cite{marek3,def_zircon,def_mono_zirc,Foster_2002_2}, an interstitial oxygen atom forms a dumbell type 
configuration in silica, zircon, zirconia and hafnia where an oxygen atom 
effectively incorporates into the Me-O-Me bond (Me is Si, Zr or Hf ion) forming a
Me-O-O-Me configuration with a O-O covalent bond (1.5 \AA\ in HfO$_2$).

Second, it follows from the calculations that both N$_2$ and oxygen interstitials may 
accept charge states 0,-1,-2 with respect to the lattice (Table \ref{tab_aff}). 
Their likely association product, the N$_2$O interstitial, is stable only in 
charge states (+2,+1,0). Next, it follows
from the electron exchange reactions shown in Table \ref{tab_elex}.11-13 that oxygen 
interstitials are more electronegative than nitrogen molecules. Therefore, relevant
reactions involving N$_2$ and O must have nitrogen molecules in the charge state
not more negative than that of the oxygen.  
Disregarding also reactions involving unstable initial or final charge states
leaves only one possible association reaction (Table \ref{tab_NO}.5), 
which is endothermic. Therefore, the formation of neutral N$_2$O interstitials
is unfavorable. Similarly, it is kinetically unlikely that interstitial NO is formed 
from N$_2$ molecules, since all reactions involving  N$_2$ dissociation 
(e.g. \ref{tab_NO}.6) are strongly endothermic. 
The fact that the N$_2$ molecules do not bond to the interstitial oxygen atoms suggests
that molecular nitrogen is ineffective in inhibiting diffusion of highly mobile oxygen 
interstitials \cite{foster02_1} and in preventing them from migration towards the 
interface.  This finding is consistent with experimental evidence 
that oxygen abundant deposition techniques lead to the formation of the SiO$_2$ 
buffer layer at the HfO$_2$/Si interface, and that the N$_2$ PDA does not 
significantly affect its thickness  \cite{Carter_aps2003,Kerber_ieee2003_1}.

\begin{table} 
\squeezetable
\caption{Energetics of the association reactions of nitrogen and the
interstitial oxygen  in m-HfO$_2$.  $E>0$ denotes an exothermic
reaction along the  arrow.}
\begin{ruledtabular} 
\begin{tabular}{llr} 
No & Reaction & Energy (eV)  \\ \hline   
1 & $  2O^0 \Rightarrow O_2^0 $            & -1.0	 \\  
2 & $ O^0 + O^- \Rightarrow O_2^- $        & -0.3	 \\  
3 & $ 2O^- \Rightarrow O_2^{--} $          & 0.7	 \\ 
4 & $ O^0 + O^{--} \Rightarrow O_2^{--} $  & 0.0 \\  
\hline 
5 & $(N_{2}) + O^0 \Rightarrow (N_2O) $ & -0.9  	\\  
6 & $(N_{2}) + O^0 \Rightarrow (NO) + N $ & -2.1 	\\ \hline 
7 & $N^0 + O^0 \Rightarrow (NO)       $ & 1.0  	        \\
8 & $N^- + O^0 \Rightarrow (NO)^-       $ & 1.2  	\\
9 & $N^0 + O^- \Rightarrow (NO)^-       $ & 1.6  	\\
10 & $N^+ + O^0 \Rightarrow (NO)^+       $ & 1.5  	\\
\end{tabular} 
\end{ruledtabular} 
\label{tab_NO} 
\end{table} 

\subsubsection{N$_2$ reactions with oxygen vacancies}

In the case of substantial oxygen vacancies concentration in the film, the N$_2$ 
interstitial species can interact with the vacant oxygen sites and  
passivate them. Based on the previous calculations 
\cite{Foster_2002_2,nano_book_2003}, we assume three-coordinated oxygen vacancies 
to dominate. We consider the stability of nitrogenous species in the three-coordinated anion lattice sites with respect to the infinitely separated 
interstitial species and the oxygen vacancy. Furthermore, we 
recall that nitrogen interstitial molecule is more electronegative than the 
oxygen vacancy (see Table \ref{tab_elex}) and discuss reactions involving
doubly charged vacancies ($V^{++}$) as the most thermodynamically probable. 
The corresponding reaction energies are summarized in Table 
\ref{tab_int2vac}, where the subscript $O$ denotes the specie incorporated into 
the oxygen vacancy. Note, that the superscript denoting the charge state corresponds to
the charge {\it with respect to the lattice}, which differs from the molecular charge
in case of the substitute molecules. For example,
the charge localized at the N$_2$ molecule  (nuclear plus electron) in the rhs of 
the reaction  (\ref{tab_int2vac}.1) is -2, so this defect is isovalent to the host 
O$^{--}$ ion, and thus, is {\it neutral} with respect to the lattice. Similarly 
$(N_2)^{+}_{O}$ in the rhs in (\ref{tab_int2vac}.2) denotes the N$_2^-$ substitution, 
which is in a charge state +1 with respect to the lattice.

\begin{table} 
\squeezetable
\caption{Energetics of the reactions of some interstitials with a
neutral 3-fold coordinated oxygen vacancy. The subscript $O$  
denotes the specie incorporated into the oxygen vacancy.
Energies are calculated in the 
assumption of the infinitely separated vacancy and interstitial
in the initial (left hand side) state. $E>0$ denotes an exothermic reaction along the arrow.}
\begin{ruledtabular} 
\begin{tabular}{llr} 
No & Reaction & Energy  \\ \hline   
1 & $ N_2^{--}+ V^{++}  \Rightarrow (N_2)^{0}_{O}   $ & 5.1   \\ 
2 & $ N_2^{-} + V^{++}  \Rightarrow (N_2)^{+}_{O}   $ & 4.7   \\
3 & $ N_2^{--} + V^{+}  \Rightarrow (N_2)^{-}_{O}   $ & 5.4   \\
4 & $ N^-    + V^{++}     \Rightarrow N^+_{O}     $ & 2.2   \\ 
5 & $ N^-    + V^{+}     \Rightarrow N^0_{O}     $ & 4.3   \\ 
6 & $ N^0    + V^{+}     \Rightarrow N^+_{O}     $ & 3.3   \\ 
7 & $ N^0    + V^0     \Rightarrow N^0_{O}     $ & 5.7   \\ 
8 & $ N^{-}  + V^0     \Rightarrow N^{-}_{O}	$ & 6.1   \\ 
9 & $ (NH_2)^{+} + V^0     \Rightarrow (NH_2)^{+}_{O} $ & 6.9   \\
10 & $ (NH)^{0} +   V^0     \Rightarrow (NH)^{0}_{O}  $ & 6.6  \\ 
11 & $ (H)^{+}  +   V^0     \Rightarrow H^{+}_{O}   $ & 2.7 \\
\end{tabular} 
\end{ruledtabular} 
\label{tab_int2vac} 
\end{table} 

One can see that most of the stable interstitial species react with the 
pre-existed anion vacancies with a very substantial energy gain. Reactions 
\ref{tab_int2vac}.1,2 correspond to a formation of N$_2$ molecule at the oxygen site, 
which is structurally similar to the O-N complex formed by the nitrogen atomic 
interstitial (see below and Fig. \ref{charge_dens}(b)). 
It is interesting to note, that the (N$_2$)$_O$ is stable in three charge
states +1,0,-1. It is evident that formation of the negative (N$_2$)$_O$ 
substitution involves either pre-existing neutral vacancies (vacancies with 2 
trapped electrons), or an electron(s) trapping by the neutral substitution. 
Although such processes are even more energetically favorable 
(e.g. \ref{tab_int2vac}.3), they might be kinetically inefficient due to the much 
lower mobility of the  V$^0$ and V$^+$ as compared to V$^{++}$.

Adding/removing an electron to/from the (N$_2$)$_O$ just 
increases/decreases the N-N bond length and decreases/increases the nearest 
neighbor Hf-N bonds. The electron affinity of N$_2$ at the oxygen vacancy is 
similar to that of the interstitial molecule (Table \ref{tab_aff}). 

The energy gain in the interstitial-vacancy reactions shown in Table \ref{tab_int2vac}
may be compared with the formation energy of the oxygen Frenkel pair in different 
charge states (8.1, 7.0 and 5.6 eV for the (V$^0$ - O$^0$), (V$^+$ - O$^-$), and
(V$^{++}$ - O${^--}$) pairs respectively \cite{Foster_2002_2}). It follows that, although 
the substitution of site oxygen ions by nitrogen molecules is endothermic, the 
substitution energies can be as low as 0.4 eV (as for N$_2^{--}$ molecular ions), so the 
processes involving oxygen replacement by nitrogen are possible at elevated temperatures.

We note that the reactions involving different charge states of the vacancy can 
be readily calculated but they are likely to be even more exothermic due to a less 
stable left hand side (e.g. reaction (\ref{tab_int2vac}.3).

\subsection{Atomic Nitrogen}

As already mentioned, a dissociation of the N$_2$ molecule in the gas phase 
costs about 10 eV. However, this energy is greatly reduced in the bulk of hafnia 
crystal. Table \ref{tab_diss} summarizes all possible reactions which would lead 
to the dissociation of N$_2$. In reactions (\ref{tab_diss}.1,2) we see that dissociation 
of the neutral molecule within the crystal is still endothermic, requiring about 3 eV 
for both neutral and charged products. The dissociation energy of the N$_2^{2-}$ 
molecular ion, although smaller, is still $\sim$1.6 eV (\ref{tab_diss}.3,4). 
Thus, even without considering the reaction barriers, the dissociation of the 
nitrogen molecule seems highly unlikely even at elevated temperatures. We also 
find that, similarly to the N$_2$ interstitial species, the N$_2$ species 
trapped by oxygen vacancies are stable with respect to dissociation. The 
corresponding reactions (\ref{tab_diss}.5-11) refer to N$_2$ species in the oxygen 
vacancy on the left and to one of the product species in the vacancy and another 
in the interstitial position on the right. As one can see, all dissociation reactions 
considered in Tab. \ref{tab_diss} are endothermic, hence, trapping of N$_2$ in 
the oxygen vacancy does not significantly reduce its dissociation energy. From this, 
one may infer that after N$_2$ PDA, the concentration of atomic or ionic 
nitrogen in the bulk film will be negligible.

\begin{table}[htbp]
\squeezetable
\caption{Dissociation reactions and associated energies for N$_{2}$ in
hafnia. The energies are calculated in the 
assumption of the infinitely separated products.
Negative energies indicate endothermic reaction in arrow direction.}
\begin{ruledtabular} 
\begin{tabular}{ccc}
No.& Reaction& Energy (eV)\tabularnewline \hline 
1& $N_{2}^{0}\Rightarrow2N^{0}$& -3.1\tabularnewline 
2& $N_{2}^{0}\Rightarrow N^{+}+N^{-}$& -3.1\tabularnewline 
3& $N_{2}^{-}\Rightarrow N^{0}+N^{-}$& -2.2\tabularnewline
4& $N_{2}^{2-}\Rightarrow N^{-}+N^{-}$& -1.1\tabularnewline \hline 
5&$\left(N_{2}\right)^{0}_{O} \Rightarrow (N^{-})_{O}+ N^{+}$ & -2.9
\tabularnewline 6& $\left(N_{2}\right)^{0}_{O}
\Rightarrow\left(N^{0}\right)_{O}+N^{0}$& -3.4\tabularnewline
7& $\left(N_{2}\right)^{0}_{O}
\Rightarrow\left(N^{+}\right)_{O}+N^{-}$& -4.4\tabularnewline
8& $\left(N_{2}\right)^{-}_{O}
\Rightarrow\left(N^{-}\right)_{O}+N^{0}$& -2.1\tabularnewline
9& $\left(N_{2}\right)^{-}_{O}
\Rightarrow\left(N^{0}\right)_{O}+N^{-}$& -2.5\tabularnewline
10& $\left(N_{2}\right)^{+}_{O}
\Rightarrow\left(N^{+}\right)_{O}+N^{0}$& -4.7\tabularnewline
11& $\left(N_{2}\right)^{+}_{O}
\Rightarrow\left(N^{0}\right)_{O}+N^{+}$& -3.6\tabularnewline
\end{tabular}\end{ruledtabular}
\label{tab_diss}
\end{table}

However, atomic nitrogen can be introduced into the system by the plasma 
assisted nitridation \cite{Lukovsky_ass2003}  (N$^+_2$), or using metal-
nitride precursors in the film deposition \cite{Afanasev_jap2004}. 
Therefore, we consider next the energetics of the atomic nitrogen species.

\subsubsection{N atom incorporation} 

\begin{figure}
\includegraphics[scale=0.50]{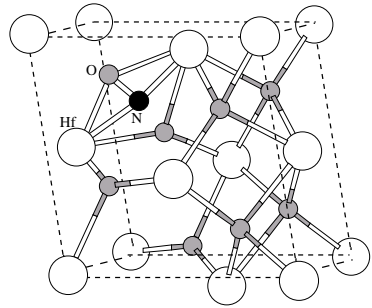}
\caption{Structure of neutral nitrogen interstitial in
hafnia. \label{n0_struct}}
\end{figure}

The interstitial nitrogen atom in its ground state configuration makes 
a covalent bond with the three-fold coordinated lattice oxygen.  This 'dumbell' 
structure is similar to that of the interstitial oxygen atom discussed 
previously in zircon \cite{def_zircon}, zirconia \cite{def_mono_zirc} and hafnia 
\cite{Foster_2002_2}. The structure of the defect is shown in Fig. 
\ref{n0_struct}, and the corresponding electron density is depicted in Fig. 
\ref{charge_dens}(b).  The electronic ground state is a doublet. This is in 
contrast with atomic nitrogen in vacuum, whose quadruplet ground state is almost 
3 eV lower than the doublet. An additional electron also fully localizes on the 
O-N bond with an affinity of 4.3 eV, increasing the bond length by about 0.1 
{\AA}.  The relaxation energy after the electron trapping is about 1.2 eV, 
compared to over 2 eV for interstitial oxygen. In contrast to the N$_2$  
molecule and an atomic oxygen, the atomic nitrogen interstitial has no affinity 
for a second electron. However, atomic nitrogen is also stable as a positive 
ion. The local relaxation in this case is distinctly different (Fig. 
\ref{charge_dens}(c)). The nitrogen-oxygen pair is now very electron deficient, 
and in fact nitrogen forms a weak bond with a second lattice oxygen
(N-O bonds are 1.46 and 1.59 \AA{}). This is 
accompanied by an energy gain of $\sim$1.2 eV. Removal of another
electron forces the nitrogen to a symmetric position, with
two strong equivalent bonds (1.40 \AA{}) to lattice oxygen sites and
an energy gain of $\sim$0.6 eV. Thus interstitial atomic nitrogen may exist 
in monoclinic hafnia in four stable charge states (+2,+1,0,-1). 

\subsubsection{N reactions with interstitial oxygen species} 
 
\begin{figure*}\includegraphics[scale=0.40]{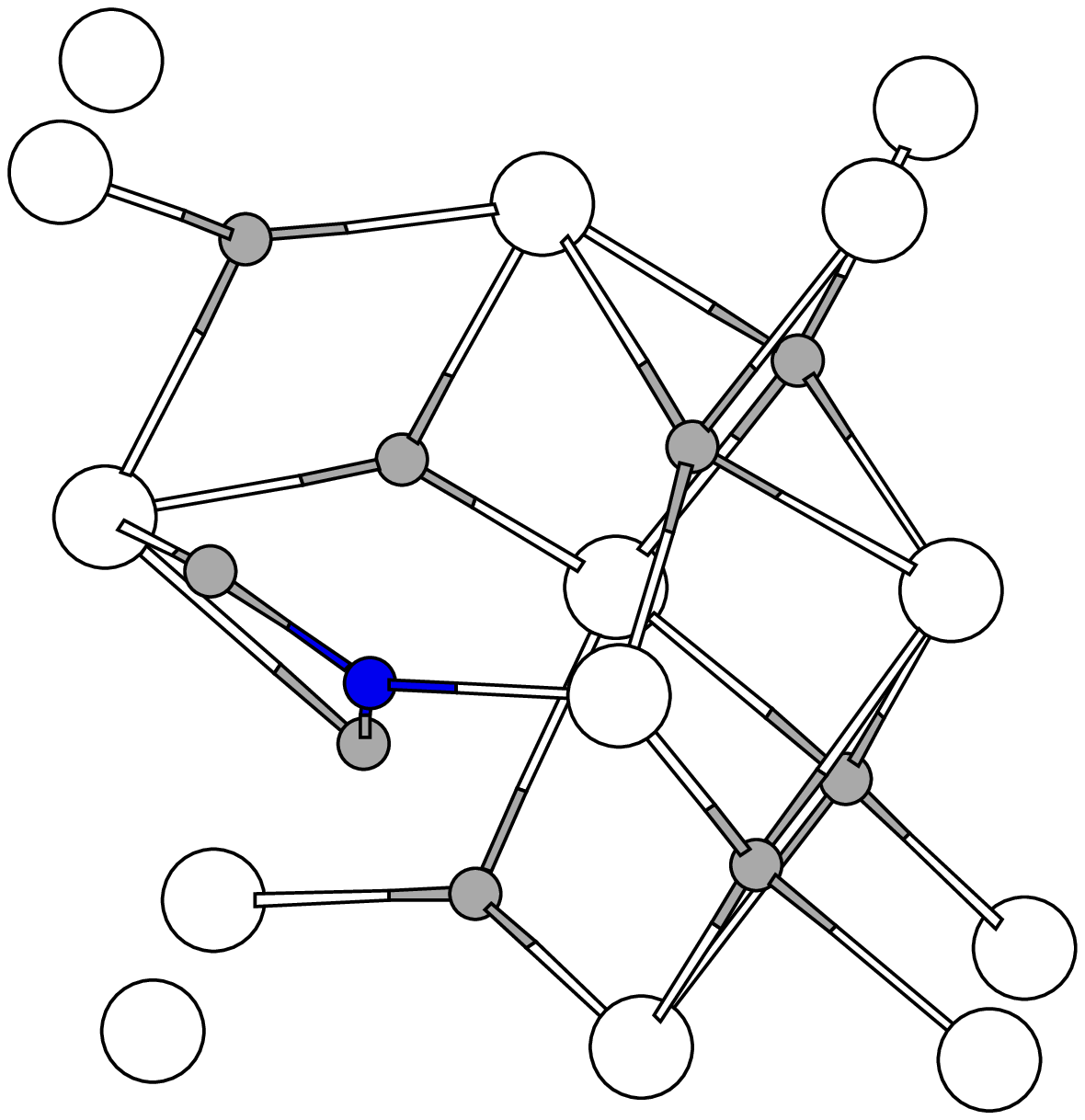}
\includegraphics[scale=0.40]{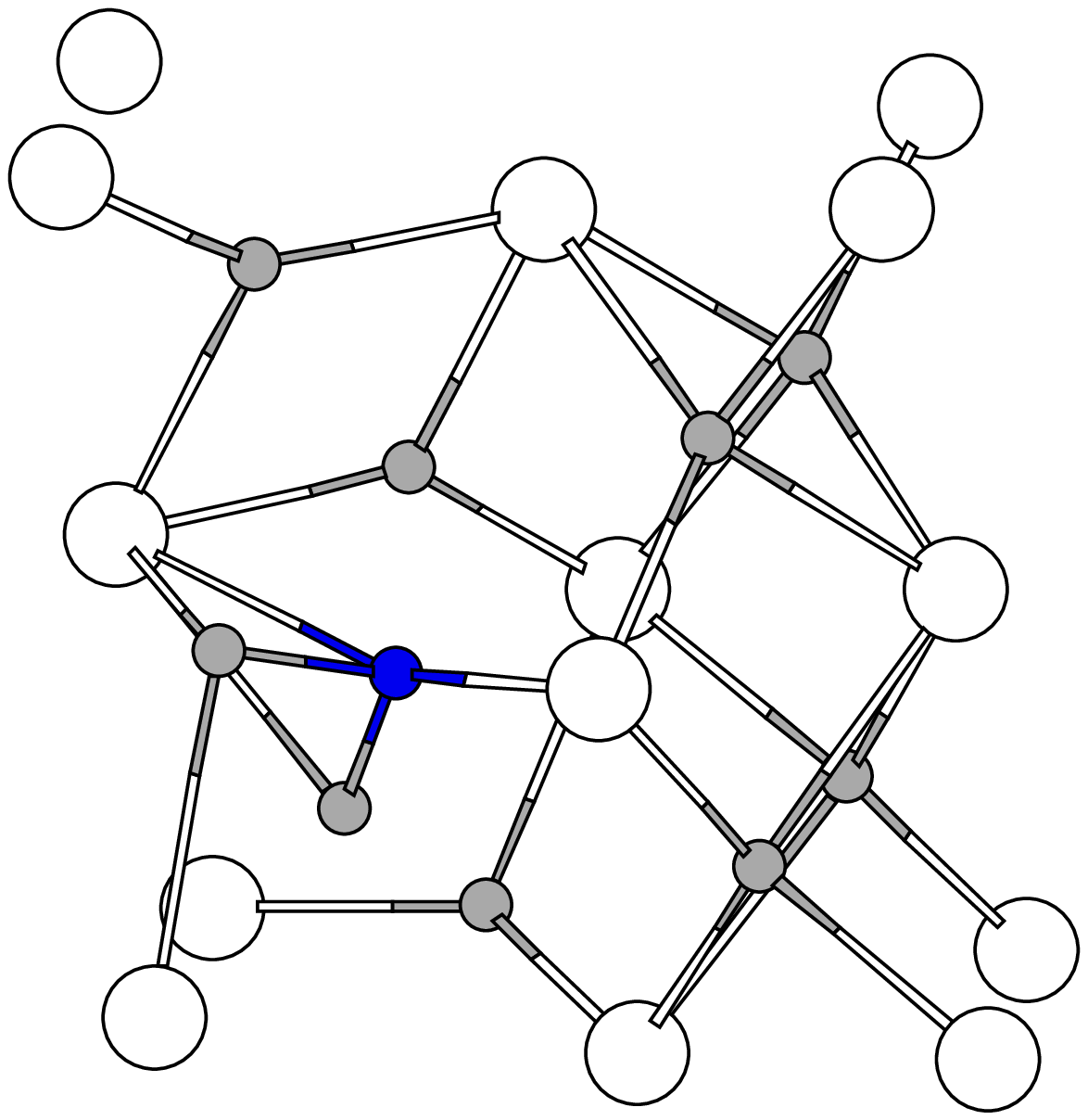}
\includegraphics[scale=0.40]{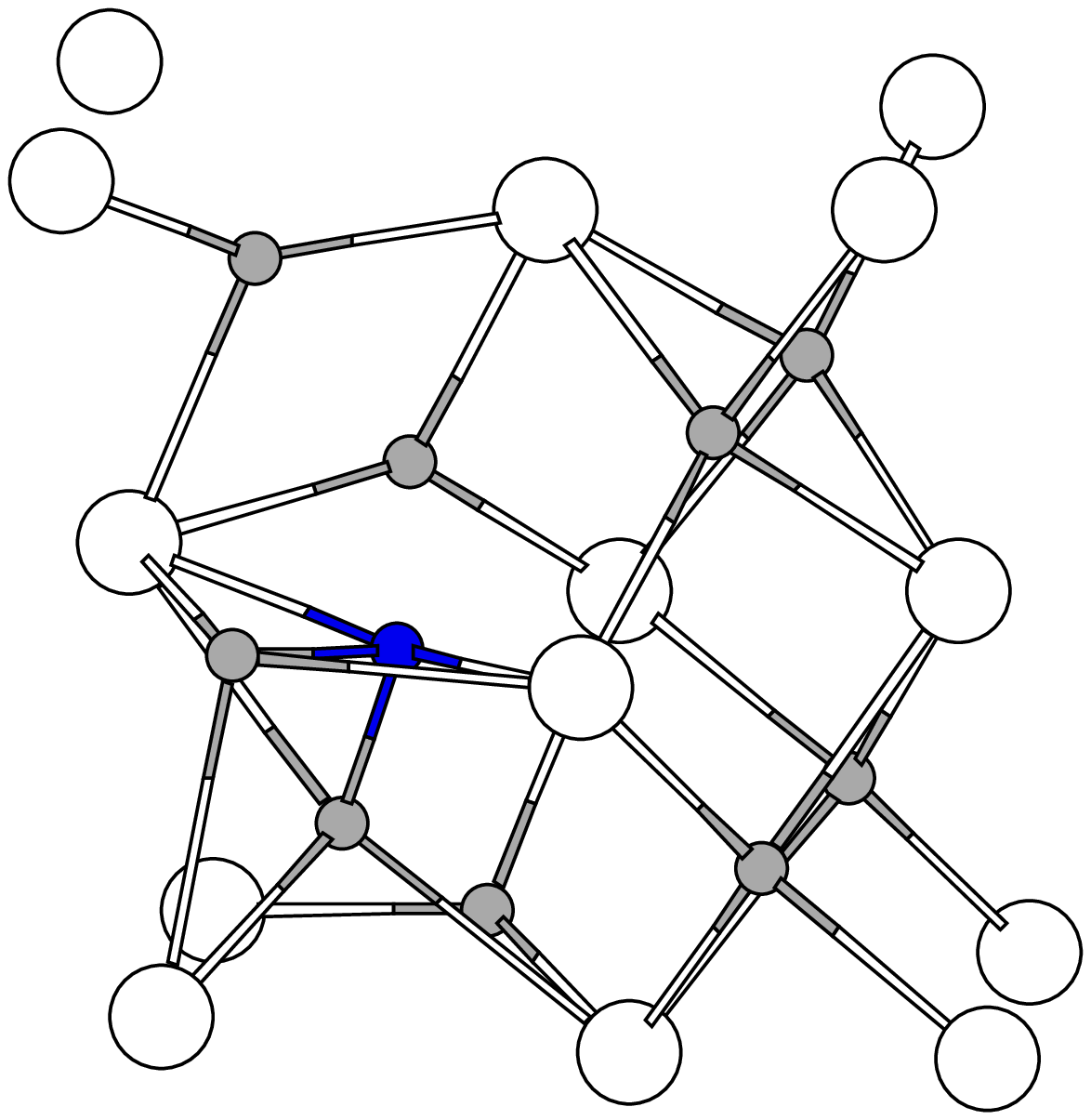}
\caption{Structure of nitric oxide interstitial molecule in
three charge states: +1 (a), 0 (b), and -1 (c). \label{NOint_struct}}
\end{figure*}

 In contrast with the N$_2$, an atomic nitrogen binds to 
an oxygen interstitial with an energy gain of around 1.0-1.5 eV 
depending on the charge state (Table \ref{tab_NO}.7-10). This process is associated 
with the breaking of the O-O bond of the 'dumbell' and formation of the NO$_2$ 
quasi-molecule (Fig. \ref{NOint_struct}b). The O-O distance is now $\sim$2.2 \AA\, 
(cf. 1.5 \AA\ in the neutral dumbell \cite{Foster_2002_2}), while the two N-O bonds 
are $\sim$1.3-1.4 \AA.  Interestingly, in this configuration there is no clear 
distinction between the interstitial and the site oxygen - the local symmetry 
for both oxygen ions remains similar. The energy gain in reactions of charged 
oxygen and nitrogen interstitial atoms is even greater (reactions 
\ref{tab_NO}.9,10). Note however that in contrast with N$_2$, the 
electronegativity of the atomic interstitial is very close to that of the oxygen
(\ref{tab_elex}.14-18), so N and O interstitials are expected to have similar 
charges.

The stability of the interstitial NO molecular species 
suggests that atomic nitrogen may play an important role in immobilizing fast 
diffusing oxygen interstitials, especially O$^-$, thereby inhibiting the growth 
of the unwanted SiO$_2$ interfacial layer during anneal.

\subsubsection{N reaction with oxygen vacancies}

In considering reactions with vacancies we recall that the interstitial atomic 
nitrogen may accept the charge states +2, +1, 0, -1, the anion vacancy - +2, +1, 0, 
and the atomic nitrogen substitution: +1,0,-1. We also take into account the result
that the nitrogen interstitial is substantially more electronegative than the vacancy,
so the relevant reactions involve nitrogen in a charge state not more positive than 
that of the vacancy.   
These considerations leave five relevant reactions shown in Table 
\ref{tab_int2vac}.4-9. Similarly to N$_2$, atomic nitrogen passivates the neutral 
vacancy with an excess energy in the range of 2-6 eV. One should also expect the
vacancy  passivation by atomic nitrogen to be more effective kinetically due to 
a higher mobility of atomic versus molecular nitrogen. Notice, however, that 
for the same reason reactions involving 
V$^{++}$ will be faster than those with V$^{+}$ or V$_{0}$ as discussed before. 

The nitrogen atom incorporation into the 
vacancy causes marginal outward relaxation of the three nearest Hf ions.  A nitrogen 
atom at a vacancy has the highest electron affinity of any of nitrogen defects 
(see Table \ref{tab_aff}). The addition of an electron (in other words N$^{-}$ 
incorporation) causes the N-Hf bond lengths to become smaller
than the O-Hf distance in the perfect lattice. The corresponding
relaxation energy  is about 0.6 eV. For N$^{+}$ substitution the
reverse relaxation occurs, i.e. the bond lengths increase, but with a similar 
energy gain.

An interesting and potentially important feature of both the nitrogen atom and 
molecule incorporated in an oxygen site is that both species can exist in the 
negative charge state. This means that the oxygen vacancy with N or N$_2$ in it 
can effectively accommodate three electrons and the third electron is strongly 
bound see (Table \ref{tab_aff}).

 On the other hand, our DFT calculations do not 
predict any electron affinity for the neutral oxygen vacancy in hafnia (see 
refs. \cite{Foster_2002_1,Foster_2002_2} for discussion). Thus the formation of 
negatively charged centers is stipulated by the large electron affinity of nitrogen 
species. Together with interstitial nitrogen species, these centers can serve as deep 
electron traps and be responsible for the negative oxide charging.

However, the existence of negatively charged nitrogen species may have also some 
positive effect as they interact with protons. For example, reaction  
\ref{tab_int_NH3}.5 clearly shows that it is energetically favorable for the 
interstitial N$^-$ to trap the proton and form a neutral NH interstitial. 
Since atomic nitrogen has a high electron affinity, and is likely to exist in 
the negative charge state, this provides a method for removing excess charge in 
the oxide due to protons - a problem often seen in experimental studies 
\cite{Haussa_jecs2002}. However, the small energy gain of -0.6 eV for the 
reaction implies the possibility of some NH dissociation at higher temperatures. 

\begin{table} 
\squeezetable
\caption{Energetics of the dissociation reactions for ammonia and its
products in m-HfO$_2$.  The stable configuration of a proton
corresponds to a bonding to the  three-coordinated oxygen. $E>0$
denotes an exothermic reaction along the  arrow.}
\begin{ruledtabular} 
\begin{tabular}{llr} 
No & Reaction & Energy (eV)  
\\ \hline   1 & $(NH_3)\Rightarrow (NH_2)^- + H^+      $ & -0.8   
\\  2 &$(NH_3) \Rightarrow (NH) + H_2          $ & -1.6    
\\ \hline   3 & $ (NH_2)^{-} \Rightarrow (NH)^0 + H^-   $ &-1.7    
\\ 4 & $ (NH_2)^{+} \Rightarrow (NH)^0 + H^+   $& -0.3    
\\ 5 & $ (NH) \Rightarrow N^-+ H^+      $ & -0.8
\\ 6 & $ (NH)^{+}\Rightarrow (N)^0 + H^+      $ & 0.1  
\end{tabular} 
\end{ruledtabular} 
\label{tab_int_NH3} 
\end{table} 

\subsection{Properties of NO and N$_2$O in hafnia}

Our calculations of the interaction of N and N$_2$ with excess oxygen in the 
oxide have established that an NO interstitial binds with lattice oxygen sites. 
Reaction \ref{tab_NO}.5 viewed from right to left clearly suggests 
that N$_2$O molecules are likely to dissociate into neutral or charged N$_2$ 
species and oxygen interstitials. It is likely that nitrogen dioxide (NO$_2$) 
will also dissociate to produce NO and oxygen interstitials. Therefore, PDA 
using nitric oxide, dinitric oxide, or nitrogen dioxide ambients will not be 
efficient in trapping oxygen and in preventing growth of SiO$_x$ buffer layer 
at the interface.

According to our calculations, the NO interstitial is a deep centre with an 
ionisation potential and electron/hole affinity similar to those of the N 
interstitial (c.f. Table \ref{tab_aff}). Being an open shell molecule, it is 
also stable in three charge states (+1,0,-1). The positive or negative 
charging of this molecule is associated with a large and not trivial lattice 
relaxation (Figure \ref{NOint_struct}). As discussed in the previous section, an 
NO interstitial bonds to the 3-coordinated lattice oxygen, forming an NO$_2^{2-
}$-like configuration at the site. The orientation of this configuration is 
strongly coupled to its charge state. The $\widehat{ONO}$ angle also changes 
from 115$^{\circ}$ for a positive molecule to 108$^{\circ}$ for the neutral, to 
103$^{\circ}$ for the negative molecule. The relaxation energy upon ionization 
is 1.3 eV, and that upon an electron trap is $\sim$1 eV, confirming the strong 
structural rearrangement. 

NO and N$_2$O molecules can interact with anion vacancies by incorporating 
with their oxygen end into the vacancy. The structure of the resulting defects 
is equivalent to that of atomic and molecular nitrogen interstitials discussed 
earlier in the paper. 


Nitrogen interstitial atoms can trap interstitial oxygen 
according to reactions \ref{tab_NO}.7-10. However, the 
contribution of this sequence of reactions is likely to be small since it 
requires an initial vacancy as a catalyst. 

\subsection{Incorporation, diffusion and dissociation of NH$_3$}

Apart from its usage for PDA of hafnia films, incorporation of ammonia into 
oxides is an interesting example of a complex molecular defect. We shall
start with the assumption that ammonia incorporates into the 
monoclinic hafnia lattice intact. However, we shall see, that such states
are metastable and more stable dissociated configurations exist. The fact that 
the molecule is hydrogen rich prompts the possibility that some hydrogen-related 
defects can be formed. In fact protons have been implicated in hole trapping in 
hafnia films and at interfaces \cite{Afanasev_jap2004}. Therefore we pay 
particular attention to possible dissociation paths of ammonia in hafnia.

\begin{figure*}
\includegraphics[scale=0.50]{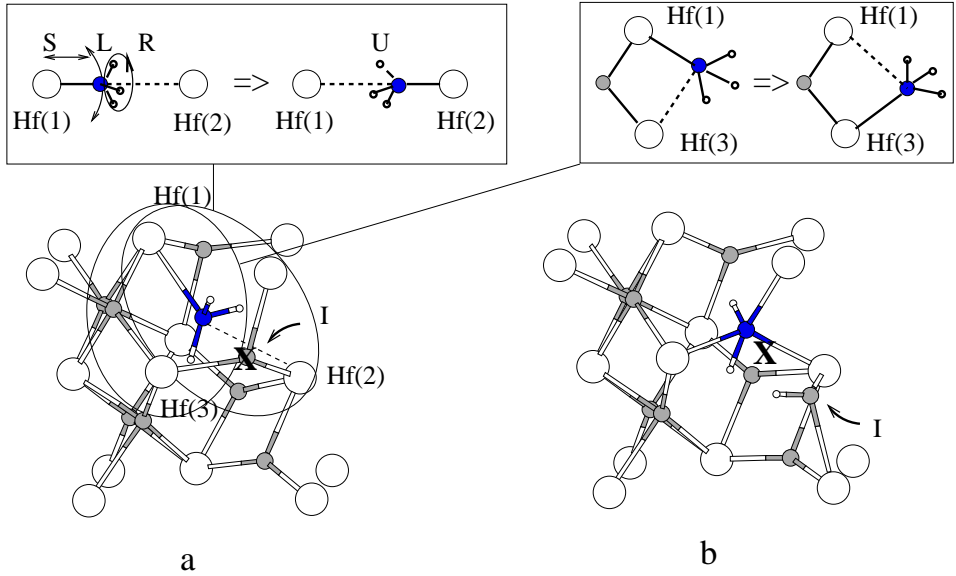}
\caption{ Equilibrium configurations for the ammonia molecule a)
intact NH$_3$ b)  locally dissociated form: NH$_2^-$ + H$^+$ (more
stable by 0.7 eV). The oxygen ion accepting the proton (marked as I)
moves off the lattice site indicated by 'X'. The insets of the figure a)
illustrate the vibrational modes involved in 'umbrella' mode assisted jump 
(left) and 'swinging' mode assisted jump (right) of the NH$_3$ molecule.}
\label{nh3_int}
\end{figure*}

In contrast to the molecular species discussed previously, ammonia is a 
nucleophilic molecule and it binds to lattice Hf ions. The stable configuration 
corresponds to the ionic Hf--N bond aligned approximately along (011)-type 
directions (Fig. ~\ref{nh3_int}(a)), with  the C$_3$ axis of the ammonia 
molecule aligned with this bond. The equilibrium Hf--N bond length is 
approximately 2.1-2.2 \AA\ . The next nearest neighbor Hf(2) ion lies 
approximately on the same axis and is separated from the nitrogen core by only 
2.5-2.8 \AA\ . 

To understand the mechanism of diffusion of ammonia inside the lattice one needs 
to analyze its vibrations. The molecule's thermal motion involves primarily four 
distinct groups of modes:  stretching of the Hf(1)--N bond (S), libration of the 
Hf(1)--N bond (L), rotation of the NH$_3$ around its $C_3$ axis (R), and the 
ammonia's so-called 'umbrella' mode (U) (Figure \ref{nh3_int}(a)). 
Two nearly degenerate L-type modes have relatively low frequencies and 
manifest themselves by large amplitude 'swings' of the NH$_3$ molecule, in which 
$C_3$ axis of the molecule remains collinear with the Hf(1)--N bond. 
These 'swings' are coupled to the R-type rotations of similar frequencies, so 
the directions of NH bonds of the molecule may adjust to a local environment. 
This L-R type thermal motion remains reversible until the C$_3$ axis is 
aligned with the Hf(1)--Hf(2) direction. Once the Hf(1)-N--Hf(2) bonds are 
aligned, the S-mode softens, so a swap between the short and long Hf-N bonds 
becomes possible. As seen in the inset of Figure \ref{nh3_int}(a),  such 
molecular hops are controlled by S and U vibrations of ammonia and  
involve only a small nitrogen displacement of under 0.4 \AA{}.  
The activation energy of the umbrella mode is $\sim$0.1 eV (gas phase data 
\cite{Manz_1997}), so frequent NH$_3$ hops between few shallow minima
are readely observed in the MD simulations even at T=300K.  An alternative 
mechanism of ammonia hops has transient states between the oxygen bridged Hf 
atoms (Hf(1) and Hf(3) in Fig. \ref{nh3_int}(a)). This involves direct S-L type 
coupling (see inset Fig. \ref{nh3_int}(a)) and has a higher activation energy. 
Despite being fairly frequent (at least one event in 1 ps at T=300K), S-U and 
S-L type hops do not contribute to molecular diffusion, since in such hops 
the molecule does not leave the original HfO$_2$ interstitial cage.

We have been unable to find a diffusion path for the whole NH$_3$ molecule.  In 
fact, further calculations show that ammonia molecule in the m-HfO$_2$  (Fig. 
\ref{nh3_int}(a)) is metastable and dissociates into NH$_2^-$ and a proton. The 
compact (NH$_2^-$-H$^+$) pair is best described as an 
NH$_2^-$ ion replacing an oxygen ion in a 3-coordinated site, while the 
displaced oxygen traps the proton and relaxes towards the volume interstitial 
position (Fig. \ref{nh3_int}(b)). The energy of this configuration is by $\sim 
0.7$ eV lower than that of the associated molecule. However, the energy barrier 
between the two configurations is substantially higher than thermal energies. 
This is reflected by the fact that the molecular NH$_3$ configuration at 300K is 
stable during the entire MD run (at least 6 ps).  However, at T = 600K, which is 
still below usual PDA temperatures, the dissociation occurs typically within 1-2 
ps. 

Stability of the dissociated configuration (Fig. \ref{nh3_int}(b)) is stipulated 
by the fact that the OH$^-$ (or NH$_2^-$) bonds are stretched by the large 
electrostatic potential gradient near the anion site. Dipole fields
of the NH$_2^-$ and OH$^-$ shearing a single site, mutually  counterballance the
crystal field effect, and stipulate stronger proton bonds and local 
neutrality. Consequently, although the initial NH$_2^-$-H$^+$ dissociation between
states {\it a} and {\it b} is exothermic, an infinite separation of H$^+$ and (NH)$^-_2$
is endothermic by 0.7 eV (Table \ref{tab_int_NH3}.1).  The latter can be 
achieved e.g. by proton diffusion away by hopping between the 3-coordinated oxygen 
ions (Fig. \ref{nh3_int}(b)). An alternative dissociation path, resulting in the 
formation of a neutral molecular hydrogen and NH (Table \ref{tab_int_NH3}.2),
is substantially less energetically favorable. 

When separated from the proton, the NH$_2^-$ radical may change its charge state 
when an external potential is applied. Our calculations suggest that this 
molecule can be either positive or negative, whilst the neutral configuration is 
thermodynamically unstable at any value of E$_F$ (Table \ref{tab_elex}.23). 
The same applies to hydrogen (\ref{tab_elex}.24), which is only stable as a proton 
or a negative ion in hafnia (see references 
\cite{VanDerWalle_nature2003,Peacock_Robertson_apl2003} for discussion on 
hydrogen). Depending on the charge state, the dissociation of NH$_2$ may lead to 
the formation of a neutral nitrohydrate and a negative or positive hydrogen ion 
(Table \ref{tab_int_NH3}.3,4). Both reactions are endothermic by 1.7 
and 0.3 eV, respectively.  Despite the apparently low dissociation energy, our 
MD simulations at T = 500 K shows that the NH$_2^+$ ion is stable - indicating 
that the dissociation barrier may be high. 

\begin{figure}

\includegraphics[scale=0.50]{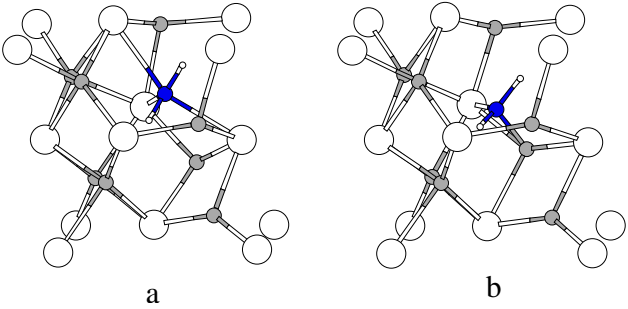}
\caption{Equilibrium configurations of the NH$_2$ interstitial (a)
NH$_2^-$; b) NH$_2^+$.}
\label{fig_2}
\end{figure}

Depending on its charge state, the NH$_2$ molecule binds to the lattice quite 
differently.  The NH$_2^-$ ion is iso-electronic to the OH$^-$ radical. 
Therefore, as expected, its most stable configuration is similar to the 
interstitial OH$^-$ centre, as reported for monoclinic zirconia 
\cite{nano_book_2003}. NH$_2^-$ binds to three Hf ions with a bond length of 
around 2.3 \AA\ (Fig. \ref{fig_2}(a)).  In contrast, the NH$_2^+$ ion makes an 
ion-covalent bond with a 3-coordinated oxygen ion  (O-N distance 1.4 \AA\ ) and 
an ionic bond with only one of the cations  (N-Hf distance 2.2 \AA)
(Fig. \ref{fig_2}(b)).

Further dissociation of NH$_2$ may lead to the formation of NH species. 
Calculations predict NH$^0$ to be stable, while NH$^-$ is unstable with respect 
to electron donation to the CB, and NH$^+$ is unstable with respect to proton
release (Table \ref{tab_int_NH3}.6). NH$^0$ molecule forms a dumbell 
configuration with a 3-coordinated lattice oxygen similar to N interstitial
(N--O distance is 1.47 \AA\ ). The equilibrium angle $\widehat{ONH}$ is 
107$^\circ$, but is soft. Finally, dissociation of NH$^0$ is energetically 
unfavorable, as discussed previously (Table \ref{tab_int_NH3}.5). 

The presented results suggest that the likely final products of ammonia PDA are 
NH$_2^-$, NH$^0$ as well as hydrogen. The latter is stable either as a proton or 
as a negative ion. The detailed dynamical behavior of hydrogen will be discussed 
elsewhere. NH$_2^-$ and NH$^0$ species are unable to trap interstitial oxygen. 
This agrees with experimental studies showing an increase in interface SiO$_2$ 
thickness during NH$_3$ PDA. In addition, NH$_2^+$ and NH$^0$ (but not NH$_2^-$) 
can react exothermally with the existing anion vacancies (see Table  
\ref{tab_int2vac}.9,10). The resultant defects create electronic levels near or 
inside the valence band of hafnia. Therefore, unlike atomic and molecular 
nitrogen species trapped in oxygen vacancy, NH$_2^-$ and NH$^0$ substitutions 
are stable in only one charge state.

\section{Discussion}

Our calculations identified the most stable nitrogen containing species and
the corresponding energy levels with respect to the band edges of the
monoclinic hafnia. It is revealed that most of the stable
nitrogen contained products (as well as intrinsic defects) may have multiple
charge states depending on the availability of the electrons in the system.
In the MOS architecture the main source of electrons in the dielectric
is the channel silicon and poly-Si or metallic gate. Therefore, to identify 
the most likely charge states, the calculated electric levels of defects must be
aligned with the silicon bands. Such an alignment is illustrated in Figure 
\ref{levels}, where the electric levels are adjusted to the experimental band
gap energy as described in the Appendix, and the experimental band off-sets are 
used for the valence and the conduction bands 
\cite{afanasev02_offs,afanasev04_offs}. Here we consider main effects caused 
by these defects.

\subsection{Structural effects}

Our calculations predict that all the neutral and negatively charged stable 
nitrogen species  (N$_2$, N, NH$_2^-$,  NH$^0$, and H$^-$) incorporate into 
oxygen vacancies with a substantial energy gain. However, this process is 
diffusion limited. Since the diffusivity of N$_2$, NH$_2$, NH$^0$, and H$^-$
is unlikely to be high, the process of vacancy passivation by molecular 
nitrogen or ammonia may be incomplete, leaving significantly non-uniform 
concentrations of interstitials and anion vacancies. In contrast, atomic 
nitrogen is expected to be rather more mobile resulting in more effective 
passivation of oxygen vacancies. A similar argument applies to the processes 
involving anion vacancies in charge states +1 and 0. They are significantly 
less mobile than the bare V$^{++}$ due to the
electrons localized inside. Consequently, the time to attain 
thermodynamic equilibrium might be long, and the subtitution by the PDA 
agent in practical anneals may be incomplete. 

\begin{figure*}
\includegraphics[scale=0.50]{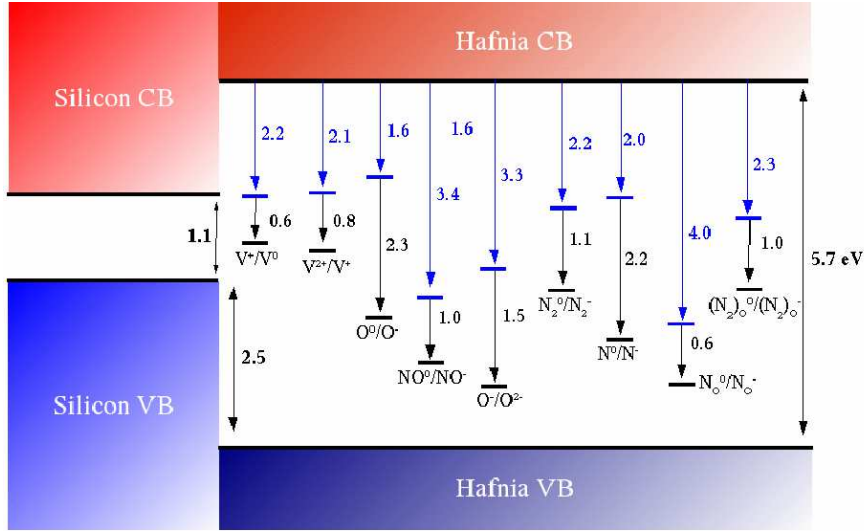}
\caption{Electron affinities of various defects in hafnia, given in
  both vertical and deeper, relaxed forms. All values are in eV.}
\label{levels}
\end{figure*}

\subsection{Electron trap effects}

Our calculations put the oxygen vacancy electrical levels into
the silicon band gap (Figure \ref{levels}), which would make them type 
{\it B} defects. However, this attribution is a result of our calculation 
procedure in which the DFT band gap correction is applied to electron 
affinities of all defects irrespectively of their origin (see Appendix). 
Such approximation may become inadequate for the doubly charged oxygen 
vacancy whose (unoccupied) electronic state splits down from the CBM due to
a strong outward displacement of the three nearest neighbour hafnium ions.
Thus, a vacancy state originates from a hafnia conduction band, and its
electron affinity ought to be calculated accordingly. Assuming that the 
vacancy level must be shifted up rigidly with CBM, one would obtain relaxed 
and unrelaxed electrical levels V$^{++}$/V$^{+}$ at ~0.5 eV and 1.3 eV 
below CBM. These values give a lower limit for the V$^{++}$ vertical electron
affinity. For the sake of consistency we retain the values obtained by the
defect independent procedure outlined in the Appendix, but acknowledge
that anion vacancies may belong to a type {\it A}, i.e be resonant with the 
Si CB. Therefore, they may act as shallow traps. Accepting this as a working 
assumption, the role of nitrogen PDA with regard to the shallow trap problem 
reduces to the effectiveness of the incorporation of nitrogen species into 
anion vacancies.

Nitrogen incorporation into the oxygen vacancy sites leads to the lowering of
the defect levels, making them less active as electron traps. In this respect,
all the considered PDA will reduce the shallow trap concentration, although to 
a different extent as discussed before. 

However, another shallow trap candidate could be a small radius electron 
polaron in the form of Hf$^{3+}$ ion, similar to the Zr$^{3+}$ centres observed
in zirconia \cite{Ryabchuk_2004}. 
The electron self-trapping if effective, may present a serious problem in the 
high-k MOSFETs.
However, our density functional calculations do not predict this defect as 
stable, possibly due to the known limitations of local approximations
regarding localisation problem \cite{hole_trapping2}. Therefore, we leave 
the problem of small polaron assisted trapping for further investigation. 

\subsection{Fixed charge effects}

Both N and N$_2$ interstitials and subsitions are deep centres, and according
to Figure \ref{levels} their relaxed affinities (electrical levels) are all
below the VBM of Si, which would make them type C defects.
This means that at equilibrium they ought to have maximum negative charge.
However, such an equilibrium state assumes that the electrons are supplied from 
the silicon valence band, which is difficult (if not impossible) for two 
reasons: 1. The effectiveness of the 
resonant tunneling strongly depends on the overlap  between the states involved 
in the process. Therefore, defects localized far away from the interface are 
ineffective electron traps, even if their energy is resonant with Si VB. 
2. Charging occurs predominantly via resonant electron tunneling from the Si VB, 
i.e. it envolves transitions on to the unoccupied defect states. These states 
(shown as unrelaxed levels in Figure \ref{levels}) are mostly of type B. Therefore, 
most of the nitrogen species are 
unavailable for a resonant tunneling from the silicon valence band. Hence, they 
may retain multiple non-equilibrium charge states for a very long time, and thus
contribute to either positive or negative charge. The possible exceptions are NO 
interstitials and N substitutions, whose electrical levels are below the Si VBM, so 
they will quickly achieve maximum negative charge.

The only definite source of positive fixed charge in our calculations are 
protons and anion vacancies. With regard to these defects, atomic 
nitrogen and molecular nitrogen will reduce vacancy concentration, but
may themselves produce negative or positive charge states as disscussed above. Ammonia
anneal will result in higher concentration of protons, but it is unclear how much of
these will be evacuated during the anneal, so the increase in positive charge
is expected with the NH$_3$ PDA. 

Finally, we would like to emphasise that high-k
materials generally favour charged defect states, due to their large polarization 
energy. This is quantified for m-HfO$_2$ in Table \ref{tab_elex}.  

We conclude that nitrogen's capacity for reducing fixed charge in hafnium 
oxide is limited. Therefore, the problem of fixed charge built up and V$_T$ 
instability is not resolved by PDA.
Hinkle and Lukowsky \cite{Lukovsky_ass2003} reported that bulk 
only nitridation does not decrease and may even increase fixed charge in the oxide 
film, which is in qualitative agreement with our calculations. 

\subsection{Growth of the interfacial SiO$_x$ layer}
 
We assume that growth of the SiO$_x$ layer at the interface is controlled by a
migration of oxygen towards the interface. Therefore, we considered 
how various PDA gases interact with oxygen interstitials. Our calculations predict
N$_2$ does not bond to oxygen interstitial in any charge state (Table \ref{tab_NO}). 
Therefore, interstitial oxygen is free to diffuse towards the interface.
In contrast, atomic N strongly bonds to the interstitial oxygen, forming less 
mobile NO$^-$ molecules.  As a result oxygen diffusion towards the interface is 
blocked, thus inhibiting growth of interfacial SiO$_x$ layer.
This conclusion is consistent with data on nitrogen 
remote plasma deposition processing \cite{Lukovsky_ass2003}, where reduction of 
interfacial silica growth has been observed.

Ammonia molecules in m-HfO$_2$ dissociate spontaneously resulting in the 
formation of NH$_2^{+/-}$, NH$^0$, H$^+$ and possibly H$_2$ and H$^-$.
Of these products, only protons will effectively bond to the interstitial
oxygen ions and exclusively to those in the charge state -2. At the same time,
the most mobile charge state of oxygen, O$^-$\cite{foster02_1}, does not 
interact strongly with protons. Therefore, ammonia PDA will not
inhibit effectively the growth of the interfacial SiO$_x$ layer. 
The same applies to the PDA with nitric oxide.

\subsection{Perspective and future work}
 
Our calculations suggest that hafnium oxynitride might have better properties as 
a gate oxide. Due to the multiplicity of the 
NO oxidation states, this material may form stable amorphous structures with 
acceptable dielectric properties. Some preliminary results on hafnium oxynitride 
gates in MOSFET have been recently reported \cite{Choi_ieee2003,Kang_apl2002},
but a more systematic study of film composition and electrical properties is 
required.

In this work we have considered only bulk structures close to their 
thermodynamic equilibrium. There exists strong experimental evidence 
\cite{Lukovsky_ass2003,Afanasev_jap2004} that in certain cases a significant 
part of the electron traps are located in the bulk of the insulator. However, 
fixed charge distant from the channel is well screened and therefore, has only 
limited effect on the channel mobility. So, to understand the processes in the 
channel, an extension of this work is required towards systems which include 
interface effects - work in this direction is currently under way. Also, we note 
that bulk DFT calculations only give an indication of the defect charge states 
that are thermodynamically possible. Some of these states may or may not be 
relevant depending on the values of the electron chemical potential. Moreover, 
charge transport through the dielectric is generally slow, so actual defect 
states may be controlled kinetically rather than thermodynamically.
Additionally, calculations of relative energies of defects in different 
charge states require empirical corrections due to various limitations of the
method applied (e.g ill-defined zero energy and spurious electrostatic
interactions in periodic boundary conditions, underestimation of the energy band 
gap {\em etc.}). Such corrections must be treated with some caution, and new 
theoretical developments are required for a more rigorous approach.

\begin{acknowledgments}
The funding of the EU Framework 5 HIKE project is gratefully 
acknowledged. This research has been supported in part by the Academy
of Finland through its Centre of Excellence Program (2000-2005). The
calculations were performed on resources provided by the HyperSpace
supercomputer Center at University College London, HPCX UK (Materials
Chemistry Consortium) and the Centre of Scientific Computing, Espoo,
Finland. We are grateful to A. M. Stoneham, V. Afanasiev, A. Asenov and A. 
Korkin for useful discussions.
\end{acknowledgments}

\appendix

\section{Calculation of Formation and Ionization Energies, and Electron and Hole 
Affinities}

The formation energy of a defect in the charge state $q$ is given by:

\begin{eqnarray}
 G(D^q,E_F) = E_D^q-E_0^q-E_D^{gas} + qE_F,
\label{eq_form}
\end{eqnarray}

\noindent
where $E_X^q$ denotes the total energy calculated for the system $X$
at the equilibrium  geometry and the excess supercell charge $q$ in
units of the electron charge $e$, $E_0^q$ is the energy of the perfect
HfO$_2$ crystal in the charge state $q$, $E_D^{gas}$ is the energy of
the isolated molecule, and all energies are calculated in the same supercell. 
Depending on the value 
of $E_F$, the thermodynamically stable charge state of the
specific defect may change.

In order to study stable charged defect states and the possible role
of defects in photo- and thermo-stimulated processes, as well as in electronic
devices one needs to know the electron affinities and ionization energies
of the defect states with respect to the bottom of the conduction band
of hafnia and to other electron or hole sources, such as silicon.To achieve 
that, we compare total energies of the initial and final
systems with the same number of electrons. The main inaccuracy of
this approach is related to the underestimation of the band-gap in the DFT 
calculations. 
This means that the defect states appear to be closer to the valence
band and conduction band edges in  hafnia. The relative error in the
energy level position with respect to the gap edges depends on a defect and is 
impossible to establish without proper calibration using experimental data.

Defining the absolute value of the defect ionization energy \( I_{p}(D^{q}) \)
as the vertical excitation energy of an electron from the defect with the
charge \( q \) to the bottom of the conduction band, we have:

\begin{equation}
\label{f5}
I_{p}(D^{q})=E^{-}_{0}+E^{q+1}_{D}-E^{0}_{0}-E^{q}_{D}+\kappa 1,
\end{equation}

where the value \( E^{q+1}_{D} \) is calculated for the geometry of the
relaxed defect with charge \( q \) and \( \kappa 1 \) is a correction
for the position of the bottom of the conduction band. Similarly we
can define the electron affinity of the defect \( \chi _{e}(D^{q}) \)
(i.e. the energy gain when the electron from the bottom of the conduction
band is trapped at the defect) as follows:

\begin{equation}
\label{f6}
\chi_{e}(D^{q})=E^{-}_{0}+E^{q}_{D}-E^{0}_{0}-E^{q-1}_{D}+\kappa 2.
\end{equation}

\noindent Here the correction \( \kappa 2 \) can be generally different
from \( \kappa 1. \) One can consider both {}``vertical'' and {}``relaxed''
electron affinities. In the latter case the lattice relaxation after
the electron trapping is included in \( E^{q-1}_{D} \). We can also
define the hole affinity of the defect \( \chi _{h}(D^{q}) \), i.e.
the energy gain when the a free hole is trapped from the top of the
valence band to the defect as follows:

\begin{equation}
\label{f7}
\chi_{h}(D^{q})=E^{+}_{0}+E^{q}_{D}-E^{0}_{0}-E^{q+1}_{D}+\kappa 3.
\end{equation}

\noindent Again, dependent on whether the lattice relaxation in the
final state is included or not, one will obtain different affinities.
The vertical hole affinity provides a useful estimate of the position
of the defect state with respect to the top of the valence band. One can also
verify, that the relaxed affinity is related to the defect's q+1/q
electric level as defined elsewhere (e.g. Ref. \cite{VandeWalle_2004}). 

To define the corrections \( \kappa 1 \), \( \kappa 2, \) \( \kappa 3 \)
we use the following considerations. i) We assume that the main inaccuracy
in defining the relative positions of defect states with respect to
the band-gap edges is due to unoccupied Kohn-Sham states, and that
the underestimated band gap is mainly due to the too low position
of the conduction band minimum. Therefore, we use an approximation
that \( \kappa 1 \) = \( \kappa 2 \) = \( \kappa  \) and \( \kappa 3 \)
= 0. These conditions are difficult to fully justify without comparison
with experiment. ii) Using these conditions and definitions
(\ref{f6}) and (\ref{f7}) it is easy to obtain:

\begin{equation}
\label{f8}
\chi _{h}(D^{q})+\chi _{e}(D^{q+1})=E_{g}(exp),
\end{equation}

\noindent where both affinities correspond to relaxed final defect
states. This condition holds in all calculations, which insures the
consistency of our approach. iii) We use the experimental value of
\( E_{g}(exp)=5.68 \) eV \cite{hf_gap} to define the difference

\begin{equation}
\label{f2}
\kappa =E_{g}(exp)-E_{g}(theor),
\end{equation}

\noindent and correct the defect ionization
potentials and electron affinities. This gives \( \kappa =5.68-3.92=1.76 \)
eV, which is used in all calculations.

Although this method is approximate, fixing the value of \( \kappa  \)
allows us to present the results of our calculations in one scale.
Another advantage is that, in order to find defect affinities with
respect to electrons at the bottom of silicon conduction band or holes
at the top 
of silicon valence band, within the same method one can
use the experimental value of band offset with Si. This scale can
be changed if a more \char`\"{}accurate\char`\"{} or relevant value
for \( \kappa  \) will be found. This will require only a shift of
our predicted values by a constant.

\bibliographystyle{apsrev} 

\bibliography{hfo2_nh3_paper,adam_refs,hafnia,tech,zirconia,silica}



\end{document}